\documentclass[conference]{IEEEtran}
\IEEEoverridecommandlockouts
\usepackage{cite}
\usepackage{amsmath,amssymb,amsfonts}
\usepackage{algorithmic}
\usepackage{graphicx}
\usepackage{textcomp}
\usepackage{xcolor}
\usepackage{booktabs}
\usepackage{multirow}
\usepackage{siunitx}
\usepackage{enumitem}
\usepackage[acronym]{glossaries}

\def\BibTeX{{\rm B\kern-.05em{\sc i\kern-.025em b}\kern-.08em
    T\kern-.1667em\lower.7ex\hbox{E}\kern-.125emX}}

\newacronym{asr}{ASR}{Automatic Speech Recognition}
\newacronym{ser}{SER}{Speech Emotion Recognition}
\newacronym{tts}{TTS}{Text-to-Speech}
\newacronym{vc}{VC}{Voice Conversion}
\newacronym{wer}{WER}{Word Error Rate}
\newacronym{cnn}{CNN}{Convolutional Neural Network}
\newacronym{auc}{AUC}{Area Under Curve}
\newacronym{roc}{ROC}{Receiver Operating Characteristics}
\newacronym{eer}{EER}{Equal Error Rate}
\newacronym{la}{LA}{Logical Access}
\newacronym{fc}{FC}{Fully Connected}

\begin{document}

\title{Comparative Analysis of ASR Methods \\for Speech Deepfake Detection\\
\thanks{This material is based on research sponsored by the Defense Advanced Research Projects Agency (DARPA) and the Air Force Research Laboratory (AFRL) under agreement number FA8750-20-2-1004. The U.S. Government is authorized to reproduce and distribute reprints for Governmental purposes notwithstanding any copyright notation thereon. The views and conclusions contained herein are those of the authors and should not be interpreted as necessarily representing the official policies or endorsements, either expressed or implied, of DARPA and AFRL or the U.S. Government.
This work was supported by the FOSTERER project, funded by the Italian Ministry of Education, University, and Research within the PRIN 2022 program.
This work was partially supported by the European Union under the Italian National Recovery and Resilience Plan (NRRP) of NextGenerationEU (PE00000014 - program ``SERICS''). This work was partially supported by the European Union - Next Generation EU under the Italian National Recovery and Resilience Plan (NRRP), Mission 4, Component 2, Investment 1.3, CUP, partnership on ``Telecommunications of the Future'' (PE00000001 - program ``RESTART'').}
}

\newcommand{\td}{$^\dagger$}
\newcommand{\tdd}{$^\ddagger$}

\author{
\parbox{0.95\linewidth}{\centering
    Davide Salvi\td, Amit Kumar Singh Yadav\tdd, Kratika Bhagtani\tdd, Viola Negroni\td,\\ Paolo Bestagini\td, Edward J. Delp\tdd
    \vspace*{0.4em}\\
    \small\centering \td Image and Sound Processing Lab (ISPL), Politecnico di Milano, Milano, Italy\\
    \small\centering \tdd  Video and Image Processing Lab (VIPER), Purdue University, West Lafayette, Indiana, USA\\
}
}

\maketitle

\begin{abstract}
Recent techniques for speech deepfake detection often rely on pre-trained self-supervised models.
These systems, initially developed for Automatic Speech Recognition (ASR), have proved their ability to offer a meaningful representation of speech signals, which can benefit various tasks, including deepfake detection.
In this context, pre-trained models serve as feature extractors and are used to extract embeddings from input speech, which are then fed to a binary speech deepfake detector.
The remarkable accuracy achieved through this approach underscores a potential relationship between ASR and speech deepfake detection.
However, this connection is not yet entirely clear, and we do not know whether improved performance in ASR corresponds to higher speech deepfake detection capabilities.
In this paper, we address this question through a systematic analysis.
We consider two different pre-trained self-supervised ASR models, Whisper and Wav2Vec 2.0, and adapt them for the speech deepfake detection task.
These models have been released in multiple versions, with increasing number of parameters and enhanced ASR performance.
We investigate whether performance improvements in ASR correlate with improvements in speech deepfake detection.
Our results provide insights into the relationship between these two tasks and offer valuable guidance for the development of more effective speech deepfake detectors.
\end{abstract}


\section{Introduction}

In recent years, techniques for creating or altering synthetic media have become increasingly widespread and accessible.
These methods can generate highly realistic synthetic content, depicting individuals in actions or behaviors that do not belong them, often making it challenging to distinguish such material from the genuine one~\cite{fakenewscientist}.
This type of manipulated data, commonly referred to as deepfakes, can lead to unpleasant consequences when used with malicious intent.
Hence, it is becoming paramount to develop tools capable of detecting it.

In the audio domain, the use of deepfakes involves the synthesis of speech using the voice of a target speaker, making them say arbitrary utterances.
This can be achieved through various techniques, either based on \gls{tts} or \gls{vc} algorithms.
To prevent potential threats arising from the malicious use of these signals, the forensic community has presented multiple approaches aimed at detecting synthetic speech~\cite{cuccovillo2022open, bhagtani2022overview}. Several detectors have been proposed~\cite{Sun_2023_CVPR, aasist, attorresi2022prosody, yang2024robust}, and extensive research have been dedicated to investigate their robustness~\cite{yadav2024fairssd, salvi2023reliability, Borzi_2022_CVPR, salvi2024listening}.

A recent trend among the latest speech deepfake detectors involves the integration of pre-trained self-supervised models.
Initially developed for the \gls{asr} task, these systems are employed as feature extractors to generate embeddings from the input speech, which are then used as input by synthetic speech detectors.
The rationale behind this approach lies in the fact that these systems have been trained on diverse speech-related tasks in a self-supervised manner, enabling them to learn robust and meaningful representations of speech signals. These representations can be effectively leveraged for various tasks beyond their original purpose, as evidenced in \gls{ser}~\cite{goron2024improving}, voice disorder detection~\cite{gupta2024addressing} and speech deepfake detection~\cite{tak22_odyssey, kawa2023improved, guo2024audio}.
This approach offers several advantages for the detection field.
First of all, leveraging pre-trained models simplifies and accelerates the training process for new detectors, as it only requires fine-tuning rather than building models from scratch.
Second, the considered \gls{asr} models have been trained on extensive and diverse datasets, different from those specifically used for speech deepfake detection. This variety helps prevent overfitting in the trained detectors and improves their generalization capabilities, which is a fundamental aspect in multimedia forensics.
Finally, these systems have demonstrated exceptional performance, often surpassing the state-of-the-art in speech deepfake detection.

The strong performance of these approaches highlights a potential relationship between \gls{asr} and speech deepfake detection. 
However, this connection is not yet entirely clear, and we do not know whether superior performance in the \gls{asr} domain translates to enhanced performance in speech deepfake detection.
In this paper, we perform a systematic analysis to answer this question.
Our approach involves adapting two well-known pre-trained \gls{asr} models for the speech deepfake detection task, i.e., Whisper and Wav2Vec 2.0.
These systems have been released in multiple versions, with increasing dimensions and progressively enhanced \gls{asr} performance.
We investigate whether performance enhancements on their original \gls{asr} task correspond to enhanced capabilities in speech deepfake detection.
Our findings provide intuitions about the relationship between \gls{asr} and detection problems, offering valuable insights for developing future detectors.

\glsreset{asr}

\section{Proposed Method}
\label{sec:method}

In this paper, we consider the problem of speech deepfake detection and investigate its relationship with the \gls{asr} task.
In particular, we build a set of speech deepfake detectors that integrate different versions of pre-trained \gls{asr} models.
Our goal is to explore whether enhancements in \gls{asr} performance correspond to improved speech deepfake detection capabilities for the overall system.

The speech deepfake detection problem is formally defined as follows.
Let us consider a discrete-time input speech signal $\mathbf{x}$ sampled with sampling frequency $f_\text{s}$, and belonging to a class $y \in \{0, 1 \}$, where \num{0} means that the signal is authentic while \num{1} indicates synthetic data.
The goal of this task is to develop a detector $\mathcal{D}$ able to estimate the class of the signal $\mathbf{x}$ as $\hat{y} \in [0,1]$, where $\hat{y}$ is an estimate of $y$ and indicates the likelihood that the signal $\mathbf{x}$ is fake.

We consider the detector $\mathcal{D}$ as logically divided into two components:
\begin{itemize}[leftmargin=*]
    \item An \textit{Embedding Extractor} $\mathcal{E}$ that maps an input speech signal $\mathbf{x}$ into an embedding $\mathbf{e}$, such as $\mathbf{e}=\mathcal{E}(\mathbf{x})$.
    \item A \textit{Deepfake Classifier} $\mathcal{C}$ that maps the embedding $\mathbf{e}$ into the deepfake detection prediction $\hat{y}$, such as $\hat{y} = \mathcal{C}(\mathbf{e})$.
\end{itemize}
In our scenario, the embedding extractor $\mathcal{E}$ corresponds to a pre-trained \gls{asr} model, while the classifier $\mathcal{C}$ includes the final classification layers.
During the training phase, we keep the parameters of $\mathcal{E}$ frozen, updating only those of $\mathcal{C}$.
This approach ensures that the results remain closely tied to the \gls{asr} performance of $\mathcal{E}$.
We explore different versions of the embedding extractor $\mathcal{E}$, each characterized by an increasing number of parameters and improved \gls{asr} performance.
Our goal is to determine whether improved \gls{asr} performance of $\mathcal{E}$ corresponds to enhanced detection capabilities for $\mathcal{D}$.

We examine two distinct \gls{asr} models as embedding extractors $\mathcal{E}$, corresponding to Whisper~\cite{radford2023robust} and Wav2Vec 2.0~\cite{baevski2020wav2vec}.
We do so since considering two distinct systems allows us to verify whether our findings are specific to a particular model or can be generalized across different systems.

Whisper~\cite{radford2023robust} is an encoder-decoder transformer model trained across several speech processing tasks, making it a general-purpose speech recognition model.
These include multilingual speech recognition, speech translation, spoken language identification, and voice activity detection.
This model is known for its noise robustness and adaptability, which allows it to maintain strong performance even in challenging acoustic conditions.
This system is available in five different sizes: tiny, base, small, medium, and large.

Wav2Vec 2.0~\cite{baevski2020wav2vec} is a neural network architecture that comprises a multi-layer~\gls{cnn} and a transformer network, trained with masked prediction tasks. It uses self-supervision to learn contextualized speech representations from the input speech waveform and has been trained considering a contrastive loss for the \gls{asr} task, minimizing the need for labeled data.
This model has been released in three different versions with varying network sizes and training strategies: base, large, and xls-r~\cite{conneau2020unsupervised}.
The base and large models were trained exclusively on English data, whereas the xls-r model, which shares the same dimensions as the large model, was trained on multilingual data.

\begin{figure}
    \centering
    \includegraphics[width=0.9\columnwidth]{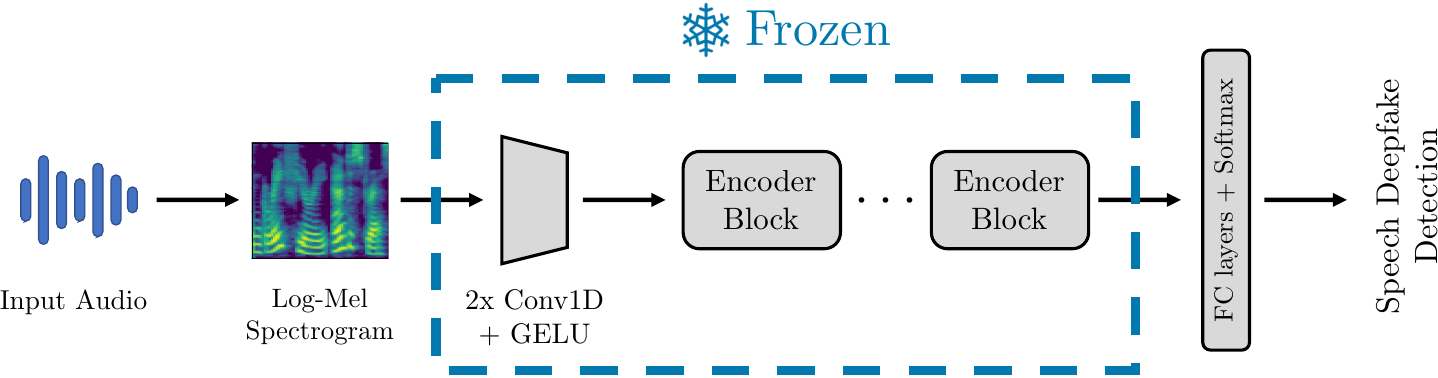}
    \caption{Whisper-based model for Speech Deepfake Detection.}
    \label{fig:whisper_pipeline}
\end{figure}

\begin{figure}
    \centering
    \includegraphics[width=0.9\columnwidth]{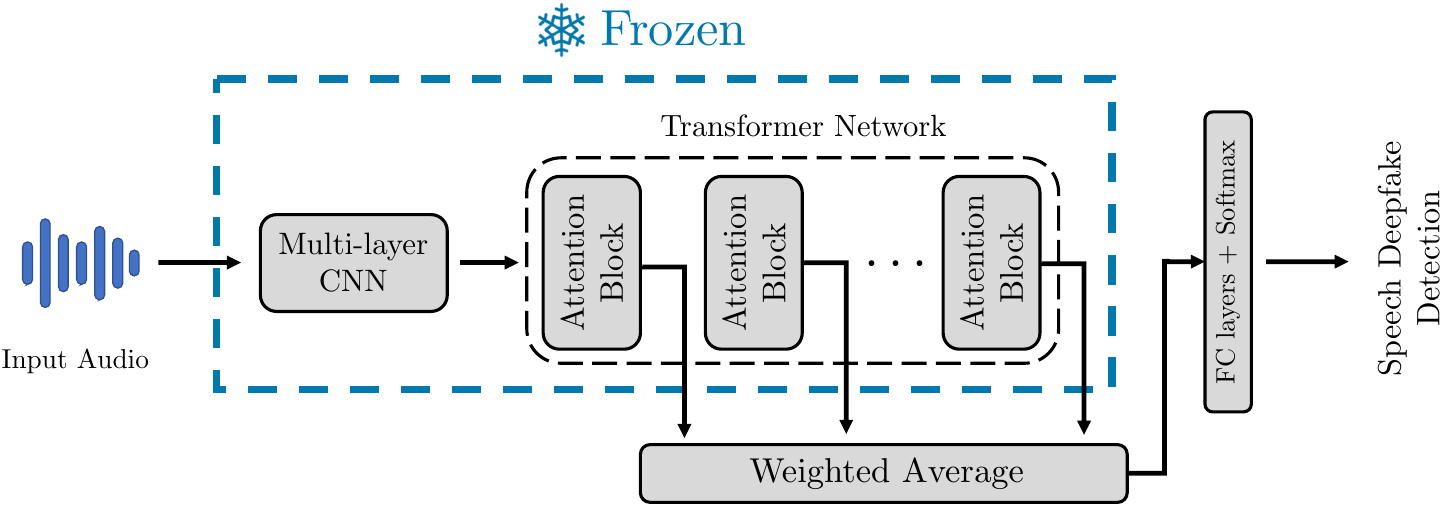}
    \caption{Wav2Vec 2.0-based model for Speech Deepfake Detection.}
    \label{fig:wav2vec_pipeline}
    \vspace{-1em}
\end{figure}

Figure~\ref{fig:whisper_pipeline} and Figure~\ref{fig:wav2vec_pipeline} show the proposed training pipeline for the two models.
As mentioned above, the parameters of the pre-trained models $\mathcal{E}$ remain frozen, and we only update the final classifier $\mathcal{C}$, which consists of a \gls{fc} network.

\section{Experimental Setup}
\label{sec:setup}

In this section, we present the evaluation setup utilized in
our experiments. We start by introducing the datasets that we considered during the training and test phases. Then, we detail the parameters that we used to train the developed detectors and the architectures of their trainable modules.

\subsection{Considered datasets}
In the experimental phase, the primary dataset we considered is ASVspoof 2019~\cite{todisco2019asvspoof}, a speech audio set created to develop antispoofing techniques for automatic speaker verification. We consider the \gls{la} partition of the dataset, which contains both real and speech deepfake data generated with various synthesis techniques.
The \textit{train} and \textit{dev} partitions include authentic signals along with synthetic speech samples generated with \num{6} different algorithms (named A01, A02, ..., A06), while the \textit{eval} partition comprises real signals and forged samples from \num{13} other algorithms (A07, ..., A19), enabling to perform open-set analyses.
We consider the \textit{train} and \textit{dev} partitions for model training and validation, respectively, and the \textit{eval} partition to test their performance.

In order to evaluate the generalization capabilities of the considered models, we also evaluate them on additional state-of-the-art datasets.
This approach ensures that our findings are not limited to a single dataset, and we can extend them to more realistic, \textit{in the wild} conditions. This is a crucial aspect in multimedia forensics, where we want our detectors to be as performant as possible, even in conditions that are different from the ones seen during training.
The additional datasets we consider are:
\begin{itemize}[leftmargin=*]
\item \emph{ASVspoof 2021}~\cite{yamagishi2021asvspoof}. It is an updated version of the 2019 dataset, where speech data are processed with different lossy codecs typically used for media storage, introducing distortions that depend on the codec type and its configuration. We considered the DeepFake (DF) partition of this set.

\item \emph{AISEC ``In-the-Wild''}~\cite{muller2022does}. It is a synthetic speech dataset that aims to replicate speech data acquired in real-world conditions. It includes almost \num{38} hours of audio clips that are either fake or real, featuring English-speaking celebrities and politicians, both from present and past.
      
\item \emph{TIMIT-TTS}~\cite{salvi2023timit}. It is a synthetic speech dataset created based on the VidTIMIT corpus considering various \gls{tts} techniques. We focus on the \textit{single speaker} partition of this dataset, which includes fake speech generated using \num{10} different \gls{tts} generators, all emulating the voice of the LJspeech dataset.
 
\item \emph{LJspeech}~\cite{ljspeech}. It is a real speech dataset containing audio clips of a single speaker reciting pieces from non-fiction books. For our experiments, we consider this set to be the authentic counterpart of TIMIT-TTS, and we consider the two sets together as a single corpus. We do so since both datasets feature speech tracks from the same speaker, making it challenging to discriminate between real and fake versions its voice.

\item \emph{FakeOrReal}~\cite{reimao2019dataset}. It is a speech dataset that includes both real and synthetic tracks. The synthetic data are generated from \gls{tts} methods, including both open-source systems and commercial tools, while the real data are collected from a series of open-source speech datasets and other sources. We considered the {\tt for-original} partition of this set.
\end{itemize}

\subsection{Training strategy}

For all the versions of both \gls{asr} methods, i.e., Whisper and Wav2Vec 2.0, we consider the same training pipeline to ensure that the differences in performance are due solely to the initial pre-trained weights.
We train the models for \num{100} epochs with an early stopping of \num{10} epochs, considering a batch size of \num{64} samples and a learning rate of $lr=10^{-4}$. We assumed Cross Entropy as the loss function and AdamW as the optimizer.
During training, we ensured that each batch was balanced by containing the same number of samples for the two classes. We did so to address the significant class imbalance between real and fake samples present in the ASVspoof 2019 dataset.
We process the audio considering a sampling frequency $f_\text{s} = $\SI{16}{\kilo\hertz}, and duration of \SI{5}{\second}.
The audio signals shorter than this duration were repeated until the desired length was reached.

As discussed in Section~\ref{sec:method}, we train only the classifier component $\mathcal{C}$ of the networks, corresponding to the final dense layers, to maintain the information provided by the pre-trained models.
Regarding Whisper, the classifier $\mathcal{C}$ consists of two fully connected layers followed by a softmax output for the final prediction. The embedding $\mathbf{e}$ extracted from the model $\mathcal{E}$ is a 2D matrix of shape (\num{1500}, $X$), where $X \in [384, 512, 768, 1024, 1280]$ for the tiny, base, small, medium, and large models, respectively.
An average pooling layer is inserted between the two dense layers to reduce the dimensionality of the input to 1D.
For Wav2Vec 2.0, the embedding $\mathbf{e}$ is obtained as a weighted sum of the outputs of all hidden layers in the pre-trained model $\mathcal{E}$.
The weights of the sum are fine-tuned during training.
The resulting embedding $\mathbf{e}$ is a 2D matrix of shape (\num{249}, $Y$), where $Y \in [768, 1024, 1024]$ for the base, large, and xls-r models, respectively.
The classifier $\mathcal{C}$ consists of three fully connected layers followed by a sigmoid output for the final prediction.
The dimensions of the layers are 1024, 128, and 1, respectively, with Rectified Linear Unit (ReLU) activation applied between them.
\section{Results}
\label{sec:results}

In this section, we evaluate and discuss the performance of the considered \gls{asr}-based models on the speech deepfake detection task.
Our goal is to investigate whether lower \gls{wer} values in \gls{asr} correlate with higher accuracy in detecting speech deepfakes.
In doing so, we evaluate the performances of the detectors in terms of \gls{roc} curves, \gls{eer}, and \gls{auc}. Optimal performance correspond to \gls{eer}$=0\%$ and \gls{auc}$=100\%$.

\begin{figure}
    \centering
    \includegraphics[width=0.65\columnwidth]{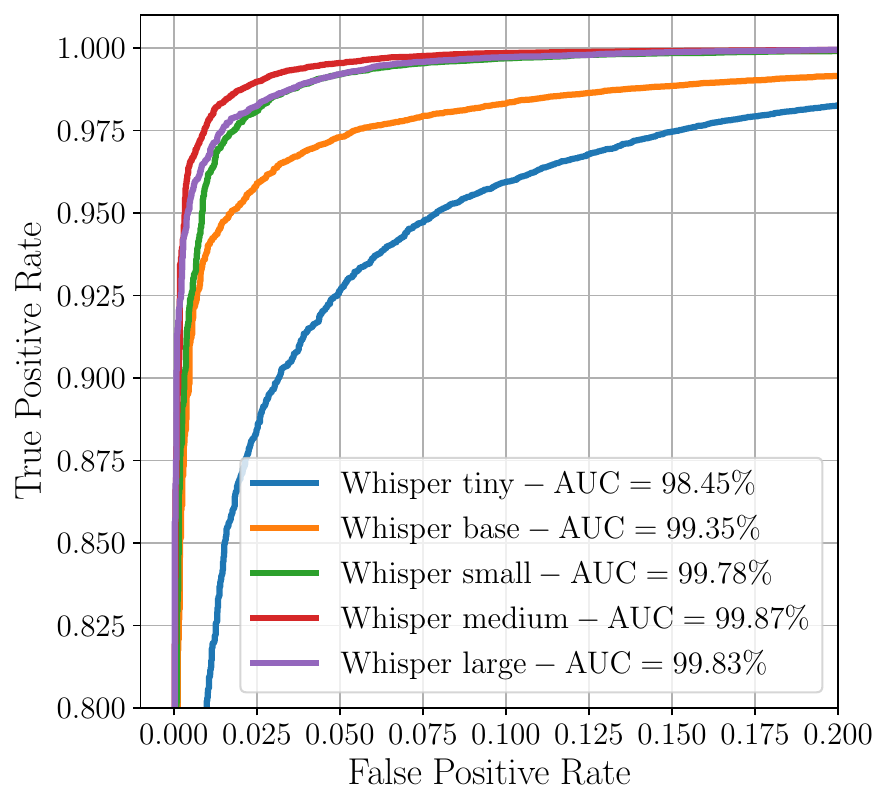}
    \caption{\gls{roc} curves of the Whisper-based models for different model sizes tested on the ASVspoof 2019 dataset. The figure's axes have been limited to improve visualization.}
    \label{fig:whisper_roc}
    \vspace{-1em}
\end{figure}

\begin{table*}
\centering
\caption{Performance of the considered models on the \gls{asr} and Speech Deepfake Detection tasks. \gls{wer} values are taken from \cite{radford2023robust}.}
\label{tab:results}
\resizebox{\textwidth}{!}{
\begin{tabular}{lcccccccccccccccccccccc}
\toprule
\multirow{2.5}{*}{Model} & \multirow{2.5}{*}{Param.}& & LibriSpeech & & \multicolumn{2}{c}{ASVspoof 2019} & & \multicolumn{2}{c}{ASVspoof 2021} & & \multicolumn{2}{c}{InTheWild} & & \multicolumn{2}{c}{TIMIT-TTS} & & \multicolumn{2}{c}{FakeOrReal} & & \multicolumn{2}{c}{Average} \\ \cmidrule{4-4} \cmidrule{6-7} \cmidrule{9-10} \cmidrule{12-13} \cmidrule{15-16} \cmidrule{18-19} \cmidrule{21-22}
                      & & & WER    & & EER    & AUC    & & EER    & AUC    & & EER    & AUC    & & EER    & AUC    & & EER             & AUC    & & EER    & AUC    \\ \midrule \midrule
Whisper tiny           & 39M & & 7.6    & & 6.21    & 98.45    & & 15.61    & 92.32    & & 37.59    & 67.37    & & 28.26  & 79.85   & & 15.33   & 92.83    & & 20.60    & 86.16             \\
Whisper base           & 74M & & 5.0    & & 3.43    & 99.35    & & 12.12    & 95.38    & & 38.34    & 65.96    & & 19.76  & 87.73   & & 8.44    & 96.81   & & 16.42    & 89.05             \\
Whisper small          & 244M & & 3.4    & & 2.12    & 99.78    & & 12.01      & \textbf{95.11}      & & 34.35    & 71.85    & & \textbf{15.53}  & \textbf{91.85}   & & \textbf{2.16}    & \textbf{99.55}   & & \textbf{13.23}    & \textbf{91.63}             \\    
Whisper medium         & 769M & & 2.9    & & \textbf{1.58}    & \textbf{99.87}    & & 12.40      & 92.32      & & 32.19    & 73.16    & & 21.74  & 87.01   & & 12.50   & 93.64    & & 16.08    & 89.20             \\
Whisper large          & 1550M & & 2.7    & & 2.00    & 99.82    & & \textbf{11.68}      & 93.82      & & \textbf{30.40}    & \textbf{77.02}    & & 23.16  & 84.70   & & 5.08    & 97.43   & & 14.46    & 90.56     \\ \midrule
Wav2Vec 2.0 base       & 95M & & 3.3    & & 3.94    & 99.34    & & 15.28      & 93.49      & & 39.83    & 64.40    & & 24.62    & 83.69    & & 19.21    & 88.69    & & 20.58    & 85.92             \\
Wav2Vec 2.0 large      & 317M & & 2.7    & & \textbf{2.54}    & \textbf{99.71}    & & \textbf{13.58}      & \textbf{94.75}      & & \textbf{28.23}    & \textbf{78.69}    & & \textbf{24.14}    & \textbf{83.92}    & & \textbf{13.07}    & \textbf{94.33}    & & \textbf{16.31}    & \textbf{90.28}             \\
Wav2Vec 2.0 xls-r      & 317M & & 4.5    & & 3.93    & 99.29    & & 17.36     & 89.12     & & 30.32     & 75.99     & & 29.07     & 75.58   & & 36.12   & 69.12   & & 23.36   & 81.82   \\ \bottomrule 
\end{tabular}
}
\vspace{-.5em}
\end{table*}

As an initial experiment, we focus on the Whisper-based models and evaluate their effectiveness on the ASVspoof2019 dataset.
Figure~\ref{fig:whisper_roc} shows the results of this analysis.
The achieved results are excellent in all the configurations, with \gls{auc} values always higher than $98\%$.
Notably, these models outperform most state-of-the-art speech deepfake detectors~\cite{muller2022does}.
We observe that, in most cases, the performance on the speech deepfake detection task improves as the dimension of the model increases. However, an exception is observed with the large model, which underperforms compared to the medium one.
This suggests the presence of a potential bottleneck, where the utility of \gls{asr} information for speech deepfake detection reaches a saturation point.

To further investigate this aspect, we extended our analysis, including also Wav2Vec 2.0 as an \gls{asr}-based detector and evaluating the models across multiple datasets unseen during training.
Table~\ref{tab:results} shows the results of this analysis.
Regarding the Wav2Vec 2.0-based models, the large model consistently outperforms the base one across all datasets.
However, the xls-r model, despite its considerable number of parameters, often performs worse than the other two.
This may be attributed to the xls-r model's multilingual training, which incorporates data from a wide variety of languages rather than focusing solely on English.
Given that all the speech deepfake datasets considered in this analysis are English-based, these conditions may not be ideal for evaluating the xls-r model.
As for the Whisper-based models, the same trend observed in the previous analysis persists: speech deepfake detection performance generally improves as \gls{asr} performance increases, but performance plateaus with the larger models.

\begin{figure}
    \centering
    \includegraphics[width=.9\columnwidth]{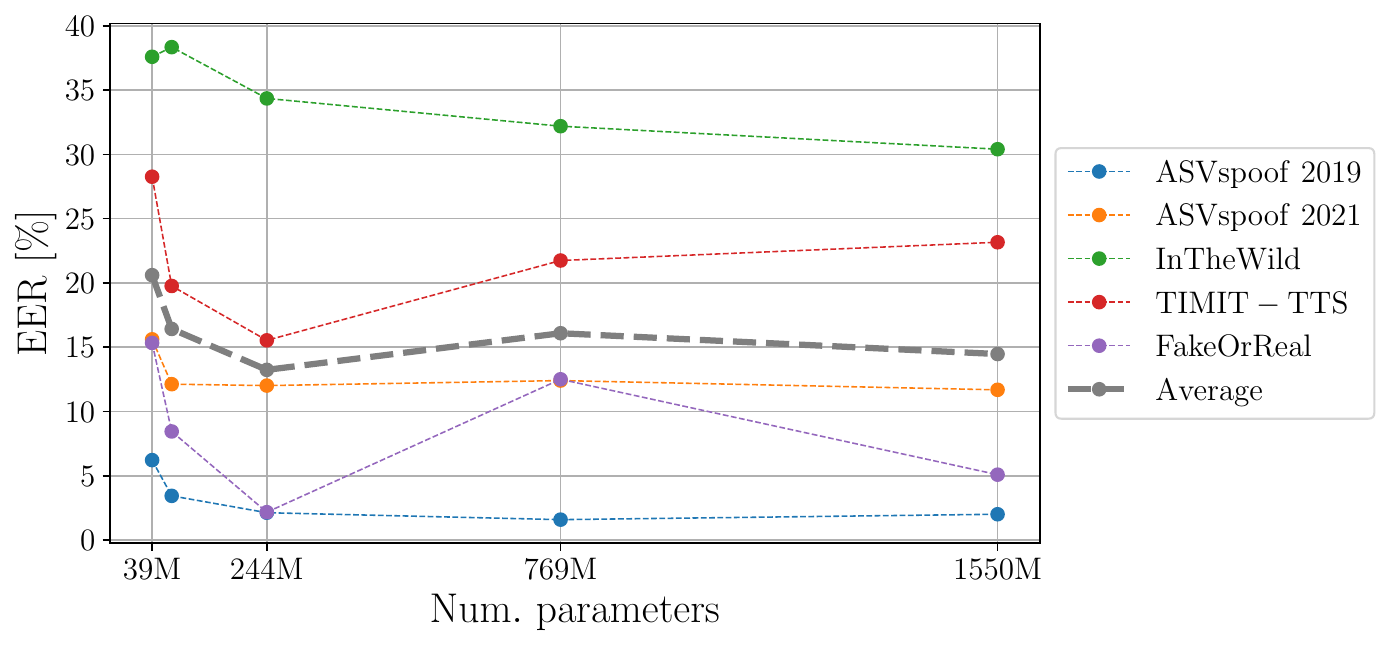}
    \caption{\gls{eer} values across the considered datasets for Whisper-based models, illustrating performance trends at increasing model sizes.}
    \label{fig:whisper_line_plot}
\end{figure}

This pattern is clearly illustrated in Figure~\ref{fig:whisper_line_plot}, which shows the \gls{eer} values across datasets as a function of model size.
On average, there is a significant improvement of nearly \num{5}\% in \gls{eer} between the tiny and base versions, with performance stabilizing thereafter.
The small model achieves the best average \gls{eer} (\num{13.23}\%), but no further improvements are observed with larger models.
This is particularly noteworthy given that the large model has six times more parameters than the small model and achieves a \num{20}\% lower \gls{wer} in the \gls{asr} task.
Despite these advantages, the large model does not yield better performance in detecting speech deepfakes.

\begin{figure}
    \centering
    \includegraphics[width=0.6\columnwidth]{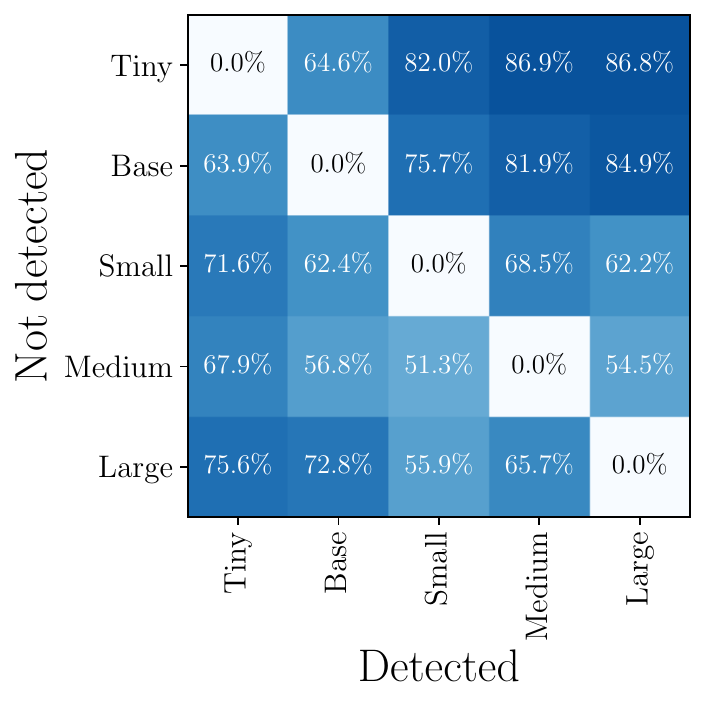}
    \caption{Detection overlap matrix showing the percentage of tracks missed by one model but detected by others.}
    \label{fig:detection_matrix}
    \vspace{-1em}
\end{figure}

As a final experiment, we further investigated the relationship between \gls{asr} and speech deepfake detection performances, exploring whether detectors based on larger \gls{asr} models can identify all the tracks detected by their smaller counterparts. This aspect is critical for understanding the extent to which these two tasks are interconnected.
Given that larger models typically provide more expressive representations of speech for the \gls{asr} task, it is reasonable to hypothesize that these richer representations would also enhance performance in speech deepfake detection. Consequently, one would expect that a larger model should be capable of detecting all the deepfake tracks correctly identified by a smaller detector, along with additional ones.
To analyze this aspect, we computed the number of tracks correctly identified by each model but misclassified by others and vice versa.
Again, we focus our analysis on the Whisper-based models, as these systems offer a greater number of dimensions.
Figure~\ref{fig:detection_matrix} shows the results of this study.
Ideally, if the \gls{asr} and speech deepfake detection tasks were perfectly aligned, the set of tracks correctly detected by a smaller model would be a strict subset of those detected by a larger model. This would result in high values in the upper triangular part of the matrix and low values in the lower triangular part.
However, the observed results deviate significantly from this ideal scenario. For instance, \num{75.6}\% of the tracks not detected by the Whisper large model are correctly identified by the tiny model, and the same consideration could be applied to several other cases.
These findings indicate that the detection capabilities of the models are less hierarchical and interconnected than initially anticipated.
It suggests that the relationship between \gls{asr} and speech deepfake detection performance are less connected than expected, with \gls{asr} performance only partially correlating with detection performance.

\section{Conclusions}
\label{sec:conclusion}

In this paper, we investigated the use of \gls{asr}-based models for speech deepfake detection, analyzing their performance across multiple datasets and configurations.

Our main goal was to determine whether enhanced performance in \gls{asr} reflects to enhanced speech deepfake detection capabilities.
We observed that while larger \gls{asr} models generally improve deepfake detection performance, this trend plateaus beyond a certain model size. Specifically, the Whisper and Wav2Vec 2.0 larger models did not consistently outperform their medium-sized counterparts in detecting speech deepfakes, suggesting the presence of a bottleneck effect where additional parameters and lower \gls{wer} do not necessarily lead to better detection performance.

Furthermore, we observed that larger models do not always capture the full range of capabilities of their smaller counterparts.
For instance, tracks misclassified by the large Whisper model were often correctly identified by the tiny model, a pattern that was also seen in other configurations.
This suggests that the \gls{asr} and speech deepfake detection tasks are only partially aligned.
Larger \gls{asr} models may not always focus on the same aspects as smaller models in performing the detection, and they may overlook subtle artifacts detectable by thinner models.

However, given the high performance obtained by all considered systems, our work emphasizes the need for further research to bridge the gap between \gls{asr} and speech deepfake detection.
Future words could focus on finding new methods to exploit the speech-related information of these systems to perform more accurate and robust speech deepfake detection.

\bibliographystyle{IEEEtran}
\bibliography{bstcontrol.bib, references.bib}

\begin{thebibliography}{10}
\providecommand{\url}[1]{#1}
\csname url@samestyle\endcsname
\providecommand{\newblock}{\relax}
\providecommand{\bibinfo}[2]{#2}
\providecommand{\BIBentrySTDinterwordspacing}{\spaceskip=0pt\relax}
\providecommand{\BIBentryALTinterwordstretchfactor}{4}
\providecommand{\BIBentryALTinterwordspacing}{\spaceskip=\fontdimen2\font plus
\BIBentryALTinterwordstretchfactor\fontdimen3\font minus \fontdimen4\font\relax}
\providecommand{\BIBforeignlanguage}[2]{{%
\expandafter\ifx\csname l@#1\endcsname\relax
\typeout{** WARNING: IEEEtran.bst: No hyphenation pattern has been}%
\typeout{** loaded for the language `#1'. Using the pattern for}%
\typeout{** the default language instead.}%
\else
\language=\csname l@#1\endcsname
\fi
#2}}
\providecommand{\BIBdecl}{\relax}
\BIBdecl

\bibitem{fakenewscientist}
S.~J. Nightingale and H.~Farid, ``{AI-synthesized faces are indistinguishable from real faces and more trustworthy},'' \emph{Proceedings of the National Academy of Sciences}, vol. 119, no.~8, p. e2120481119, 2022.

\bibitem{cuccovillo2022open}
L.~Cuccovillo, C.~Papastergiopoulos, A.~Vafeiadis, A.~Yaroshchuk, P.~Aichroth, K.~Votis, and D.~Tzovaras, ``Open challenges in synthetic speech detection,'' in \emph{IEEE International Workshop on Information Forensics and Security (WIFS)}, 2022.

\bibitem{bhagtani2022overview}
K.~Bhagtani, A.~K.~S. Yadav, E.~R. Bartusiak, Z.~Xiang, R.~Shao, S.~Baireddy, and E.~J. Delp, ``{An Overview of Recent Work in Multimedia Forensics},'' in \emph{IEEE Conference on Multimedia Information Processing and Retrieval}, 2022.

\bibitem{Sun_2023_CVPR}
C.~Sun, S.~Jia, S.~Hou, and S.~Lyu, ``{AI-Synthesized Voice Detection Using Neural Vocoder Artifacts},'' in \emph{IEEE/CVF Conference on Computer Vision and Pattern Recognition Workshop (CVPRW)}, 2023.

\bibitem{aasist}
J.-w. Jung, H.-S. Heo, H.~Tak, H.-j. Shim, J.~S. Chung, B.-J. Lee, H.-J. Yu, and N.~Evans, ``{AASIST: Audio Anti-Spoofing Using Integrated Spectro-Temporal Graph Attention Networks},'' in \emph{IEEE International Conference on Acoustics, Speech and Signal Processing (ICASSP)}, 2022.

\bibitem{attorresi2022prosody}
L.~Attorresi, D.~Salvi, C.~Borrelli, P.~Bestagini, and S.~Tubaro, ``{Combining Automatic Speaker Verification and Prosody Analysis for Synthetic Speech Detection},'' in \emph{International Conference on Pattern Recognition (ICPR)}, 2022.

\bibitem{yang2024robust}
Y.~Yang, H.~Qin, H.~Zhou, C.~Wang, T.~Guo, K.~Han, and Y.~Wang, ``A robust audio deepfake detection system via multi-view feature,'' in \emph{IEEE International Conference on Acoustics, Speech and Signal Processing (ICASSP)}.\hskip 1em plus 0.5em minus 0.4em\relax IEEE, 2024.

\bibitem{yadav2024fairssd}
A.~K.~S. Yadav, K.~Bhagtani, D.~Salvi, P.~Bestagini, and E.~J. Delp, ``Fairssd: Understanding bias in synthetic speech detectors,'' in \emph{IEEE/CVF Conference on Computer Vision and Pattern Recognition Workshop (CVPRW)}, 2024.

\bibitem{salvi2023reliability}
D.~Salvi, P.~Bestagini, and S.~Tubaro, ``Reliability estimation for synthetic speech detection,'' in \emph{IEEE International Conference on Acoustics, Speech and Signal Processing (ICASSP)}.\hskip 1em plus 0.5em minus 0.4em\relax IEEE, 2023.

\bibitem{Borzi_2022_CVPR}
S.~Borz{\`\i}, O.~Giudice, F.~Stanco, and D.~Allegra, ``{Is Synthetic Voice Detection Research Going Into the Right Direction?}'' in \emph{IEEE/CVF Conference on Computer Vision and Pattern Recognition Workshop (CVPRW)}, 2022.

\bibitem{salvi2024listening}
D.~Salvi, T.~S. Balcha, P.~Bestagini, and S.~Tubaro, ``{Listening Between the Lines: Synthetic Speech Detection Disregarding Verbal Content},'' in \emph{IEEE International Conference on Acoustics, Speech, and Signal Processing Workshops (ICASSPW)}.\hskip 1em plus 0.5em minus 0.4em\relax IEEE, 2024.

\bibitem{goron2024improving}
E.~Goron, L.~Asai, E.~Rut, and M.~Dinov, ``{Improving Domain Generalization in Speech Emotion Recognition with Whisper},'' in \emph{IEEE International Conference on Acoustics, Speech and Signal Processing (ICASSP)}.\hskip 1em plus 0.5em minus 0.4em\relax IEEE, 2024.

\bibitem{gupta2024addressing}
R.~Gupta, C.~Madill, D.~R. Gunjawate, D.~D. Nguyen, and C.~T. Jin, ``{Addressing Data Scarcity in Voice Disorder Detection with Self-Supervised Models},'' in \emph{IEEE International Conference on Acoustics, Speech and Signal Processing (ICASSP)}, 2024.

\bibitem{tak22_odyssey}
H.~Tak, M.~Todisco, X.~Wang, J.~weon Jung, J.~Yamagishi, and N.~Evans, ``{Automatic Speaker Verification Spoofing and Deepfake Detection Using Wav2vec 2.0 and Data Augmentation},'' in \emph{Speaker and Language Recognition Workshop (Odyssey)}, 2022.

\bibitem{kawa2023improved}
P.~Kawa, M.~Plata, M.~Czuba, P.~Szymanski, and P.~Syga, ``{Improved DeepFake Detection Using Whisper Features},'' in \emph{Conference of the International Speech Communication Association (INTERSPEECH)}, 2023.

\bibitem{guo2024audio}
Y.~Guo, H.~Huang, X.~Chen, H.~Zhao, and Y.~Wang, ``{Audio Deepfake Detection With Self-Supervised Wavlm And Multi-Fusion Attentive Classifier},'' in \emph{IEEE International Conference on Acoustics, Speech and Signal Processing (ICASSP)}.\hskip 1em plus 0.5em minus 0.4em\relax IEEE, 2024.

\bibitem{radford2023robust}
A.~Radford, J.~W. Kim, T.~Xu, G.~Brockman, C.~McLeavey, and I.~Sutskever, ``Robust speech recognition via large-scale weak supervision,'' in \emph{International Conference on Machine Learning (ICML)}.\hskip 1em plus 0.5em minus 0.4em\relax PMLR, 2023.

\bibitem{baevski2020wav2vec}
A.~Baevski, Y.~Zhou, A.~Mohamed, and M.~Auli, ``{wav2vec 2.0: A framework for self-supervised learning of speech representations},'' \emph{Advances in Neural Information Processing Systems (NeurIPS)}, 2020.

\bibitem{conneau2020unsupervised}
P.~Kawa, M.~Plata, M.~Czuba, P.~Szymanski, and P.~Syga, ``{Improved DeepFake Detection Using Whisper Features},'' in \emph{Conference of the International Speech Communication Association (INTERSPEECH)}, 2021.

\bibitem{todisco2019asvspoof}
M.~Todisco, X.~Wang, V.~Vestman, M.~Sahidullah, H.~Delgado, A.~Nautsch, J.~Yamagishi, N.~Evans, T.~Kinnunen, and K.~A. Lee, ``{ASVspoof 2019: Future horizons in spoofed and fake audio detection},'' in \emph{Conference of the International Speech Communication Association (INTERSPEECH)}, 2019.

\bibitem{yamagishi2021asvspoof}
J.~Yamagishi, X.~Wang, M.~Todisco, M.~Sahidullah, J.~Patino, A.~Nautsch, X.~Liu, K.~A. Lee, T.~Kinnunen, N.~Evans \emph{et~al.}, ``{ASVspoof 2021: accelerating progress in spoofed and deepfake speech detection},'' in \emph{Automatic Speaker Verification and Spoofing Countermeasures Challenge}, 2021.

\bibitem{muller2022does}
N.~M. M{\"u}ller, P.~Czempin, F.~Dieckmann, A.~Froghyar, and K.~B{\"o}ttinger, ``Does audio deepfake detection generalize?'' in \emph{Conference of the International Speech Communication Association (INTERSPEECH)}, 2022.

\bibitem{salvi2023timit}
D.~Salvi, B.~Hosler, P.~Bestagini, M.~C. Stamm, and S.~Tubaro, ``{TIMIT-TTS: a Text-to-Speech Dataset for Multimodal Synthetic Media Detection},'' \emph{IEEE Access}, 2023.

\bibitem{ljspeech}
\BIBentryALTinterwordspacing
{Keith Ito and Linda Johnson}, ``The {LJS}peech dataset,'' 2017. [Online]. Available: \url{https://keithito.com/LJ-Speech-Dataset/}
\BIBentrySTDinterwordspacing

\bibitem{reimao2019dataset}
R.~Reimao and V.~Tzerpos, ``{FoR: A dataset for synthetic speech detection},'' in \emph{International Conference on Speech Technology and Human-Computer Dialogue (SpeD)}.\hskip 1em plus 0.5em minus 0.4em\relax IEEE, 2019.

\end{thebibliography}

\end{document}